\documentclass[twocolumn,showpacs,aps]{revtex4}

\begin{document}

 \newcommand{\bq}{\begin{equation}}
 \newcommand{\eq}{\end{equation}}
 \newcommand{\bqn}{\begin{eqnarray}}
 \newcommand{\eqn}{\end{eqnarray}}
 \newcommand{\nb}{\nonumber}
 \newcommand{\lb}{\label}
\newcommand{\PRL}{Phys. Rev. Lett.}
\newcommand{\PL}{Phys. Lett.}
\newcommand{\PR}{Phys. Rev.}
\newcommand{\CQG}{Class. Quantum Grav.}

\title{Cosmological perturbations in
Horava-Lifshitz theory without detailed balance}

\author{Anzhong Wang$\,^{1}$}

\author{Roy Maartens$\,^{2}$}

\affiliation{$^{1}$GCAP-CASPER, Physics Department, Baylor
University, Waco, TX 76798-7316, USA\\
$^{2}$Institute of Cosmology \& Gravitation, University of
Portsmouth, Portsmouth PO1 3FX, UK }

\date{\today}

\begin{abstract}

In the Horava-Lifshitz theory of quantum gravity, two conditions
-- detailed balance and projectability -- are usually assumed. The
breaking of projectability simplifies the theory, but it leads to
serious problems with the theory. The breaking of detailed balance
leads to a more complicated form of the theory, but it appears to
resolve some of the problems. Sotiriou, Visser and Weinfurtner
formulated the most general theory of Horava-Lifshitz type without
detailed balance. We compute the linear scalar perturbations of
the FRW model in this form of HL theory. We show that the
higher-order curvature terms in the action lead to a gravitational
effective anisotropic stress on small scales. Specializing to a
Minkowski background, we study the spin-0 scalar mode of the
graviton, using a gauge-invariant analysis, and find that it is
stable in both the infrared and ultraviolet regimes for $0 \le \xi
\le 2/3$. However, in this parameter range the scalar mode is a
ghost.

\end{abstract}


\maketitle

\section{Introduction}
\renewcommand{\theequation}{1.\arabic{equation}} \setcounter{equation}{0}

Recently, Horava proposed a quantum gravity theory \cite{Horava},
motivated by the Lifshitz theory in solid state physics
\cite{Lifshitz}. Horava-Lifshitz (HL) theory is non-relativistic
and power-counting ultraviolet (UV)-renormalizable, and should
recover general relativity in the infrared (IR) limit. The
effective speed of light diverges in the UV regime, and this
potentially resolves the horizon problem without invoking an
inflationary scenario. In \cite{LMP,KK}, the general field
equations were derived, and given explicitly for FRW cosmology,
from which it can be seen that the spatial curvature is enhanced
by higher-order curvature terms. This could open a new approach to
the flatness problem and to a bouncing universe
\cite{calcagni,brand}. It was also shown that almost
scale-invariant super-horizon curvature perturbations can be
produced without inflation \cite{Muk}.

Horava assumed two conditions -- detailed balance and
projectability. He also considered the case without detailed
balance. So far most of the work
\cite{LMP,KK,brand,Cosmos,BHs,others} on HL theory has abandoned
projectability. However, breaking the projectability condition
seems problematic \cite{Mukb,Koch}. With detailed balance, it was
shown that matter is not UV stable \cite{calcagni}. In addition, a
non-zero negative cosmological constant is required, and it also
breaks the parity in the purely gravitational sector \cite{SVW}.

Because Lorentz invariance is broken in the UV, HL theory contains
a reduced set of diffeomorphisms, and as a result, a spin-0 mode
of the graviton appears. This mode is potentially dangerous and
may cause strong coupling problems that prevent the recovery of
general relativity in the IR limit \cite{CNPS,Cai,BPS}.

In order to avoid these problems, one possibility is to keep the
projectability condition. With this condition, Mukohyama argued
that the problems found in \cite{CNPS,BPS} can be solved by the
repulsive gravitational force due to the nonlinear higher
curvature terms \cite{Mukb}. In addition, without the
projectability condition, the theory seems to be inconsistent
\cite{LP}.

By abandoning detailed balance but still keeping the
projectability condition, Sotiriou, Visser and Weinfurtner (SVW)
showed that the most general such HL theory can be properly
formulated with eight independent coupling constants, in addition
to the Newton and cosmological ones  \cite{SVW}. Among these eight
coupling constants, one is associated with the kinetic energy,
which leads to the spin-0 scalar graviton, and the other seven are
all related to the breaking of Lorentz invariance, which are
highly suppressed by the Planck scale in the IR limit.

In this paper, we study linear scalar perturbations of FRW models
in the SVW set-up. The paper is organized as follows: In Sec. II,
we give a brief introduction to the generalized HL theory
formulated by SVW   \cite{SVW}. In particular, we add matter
fields (not considered in \cite{SVW}) and generalize the dynamical
equations and the Hamiltonian and super-momentum constraints.
Calcagni constructed the action for a scalar field
\cite{calcagni}, while Kiritsis and Kofinas considered the same
problem, and then generalized it to a vector field \cite{KK}.
However,  the general coupling of matter to this theory has not
been worked out yet, since we no longer have the guide of Lorentz
invariance. We shall not be concerned with this issue in the
present paper, and simply assume that it can be done and
represented by a general matter action, from which we can derive
the conservation laws of the matter field. In Sec. III, we present
the Friedmann-like field equations for FRW models with any
curvature $k$.  In Sec. IV, we first briefly discuss different
gauge choices, and then study the linear scalar perturbations of
FRW models, working in the quasi-longitudinal gauge. We obtain the
perturbations of the dynamical equations and the Hamiltonian and
super-momentum constraints. We also compute the perturbed matter
conservation equations. In Sec. V, we study the spin-0 scalar mode
of the graviton in a Minkowski background, by specializing the
formulas developed in Sec. IV. We find that this scalar mode,
which could potentially undermine the recovery of general
relativity in the IR limit, is in fact stable in both the IR and
UV regimes for $0 \le \xi \le 2/3$. However, it should be noted that
in this range, the kinetic term of the scalar mode in the action
has the wrong sign, as is evident in Horava's original results
\cite{Horava}, so that the mode is a ghost \cite{Mukb,BPS,KA}. In
Sec. VI, we restrict to perturbations of the flat FRW model, and
find that the corresponding field equations are considerably
simplified. In Sec. VII, we present conclusions.

\section{Horava-Lifshitz Gravity Without Detailed Balance}

\renewcommand{\theequation}{2.\arabic{equation}} \setcounter{equation}{0}

We give a very brief introduction to HL gravity without detailed
balance, but with the projectability condition. (For further
details, see \cite{SVW}.) The dynamical variables are $N, \;
N^{i}$ and $g_{ij}\; (i, \; j = 1, 2, 3)$, in terms of which the
metric takes the ADM form,
 \bq \lb{2.1}
ds^{2} = - N^{2}dt^{2} + g_{ij}\left(dx^{i} + N^{i}dt\right)
     \left(dx^{j} + N^{j}dt\right).
 \eq
The theory is invariant under the scalings
 \bqn
&& t \rightarrow {\ell}^{3} t,\;\;\; x^{i}  \rightarrow {\ell}
x^{i}\,,\nb\\ \lb{2.3}&& N \rightarrow  \ell^{-2} N ,\;  N^{i}
\rightarrow {\ell}^{-2} N^{i},\; g_{ij} \rightarrow g_{ij}.
 \eqn
The projectability condition requires a homogeneous lapse
function:
 \bq \lb{2.3a}
N = N(t), \;\;\; N^{i} =  N^{i}\left(t, x^{k}\right),\;\;\; g_{ij}
= g_{ij} \left(t, x^{k}\right).
 \eq
This is  invariant under the gauge transformations,
 \bqn
\lb{gauge} \tilde{t}  =  t +   \chi^{0}(t) , \tilde{x}^{i} = x^{i}
+  \chi^{i} \left(t, x^{k}\right).
 \eqn

The total action consists of kinetic, potential and matter parts,
 \bqn \lb{2.4}
S = \zeta^2\int dt d^{3}x N \sqrt{g} \left({\cal{L}}_{K} -
{\cal{L}}_{{V}}+\zeta^{-2} {\cal{L}}_{M} \right),
 \eqn
where $g={\rm det}\,g_{ij}$, and
 \bqn \lb{2.5}
{\cal{L}}_{K} &=& K_{ij}K^{ij} - \left(1-\xi\right)  K^{2},\nb\\
{\cal{L}}_{{V}} &=& 2\Lambda - R + \frac{1}{\zeta^{2}}
\left(g_{2}R^{2} +  g_{3}  R_{ij}R^{ij}\right)\nb\\
& & + \frac{1}{\zeta^{4}} \left(g_{4}R^{3} +  g_{5}  R\;
R_{ij}R^{ij}
+   g_{6}  R^{i}_{j} R^{j}_{k} R^{k}_{i} \right)\nb\\
& & + \frac{1}{\zeta^{4}} \left[g_{7}R\nabla^{2}R +  g_{8}
\left(\nabla_{i}R_{jk}\right)
\left(\nabla^{i}R^{jk}\right)\right].
 \eqn
Here $\zeta^{2} = 1/{16\pi G} $, the covariant derivatives and
Ricci and Riemann terms all refer to the three-metric $g_{ij}$,
and $K_{ij}$ is the extrinsic curvature,
 \bq \lb{2.6}
K_{ij} = \frac{1}{2N}\left(- \dot{g}_{ij} + \nabla_{i}N_{j} +
\nabla_{j}N_{i}\right),
 \eq
where $N_{i} = g_{ij}N^{j}$. The constants $\xi, g_{I}\,
(I=2,\dots 8)$  are coupling constants, and $\Lambda$ is the
cosmological constant. It should be noted that Horava included a
cross term $C_{ij}R^{ij}$, where $C_{ij}$ is the Cotton tensor.
This term scales as    $\ell^{5}$ and explicitly violates parity.
To restore parity,  SVW excluded this term \cite{SVW}.

In the IR limit, all the high order curvature terms (with
coefficients $g_I$) drop out, and the total action reduces when
$\xi = 0$ to the Einstein-Hilbert action.

Variation with respect to the lapse function $N(t)$ yields the
Hamiltonian constraint,
 \bq \lb{eq1}
\int{ d^{3}x\sqrt{g}\left({\cal{L}}_{K} + {\cal{L}}_{{V}}\right)}
= 8\pi G \int d^{3}x {\sqrt{g}\, J^{t}},
 \eq
where
 \bq \lb{eq1a}
J^{t} = 2\left(N\frac{\delta{\cal{L}}_{M}}{\delta N} +
{\cal{L}}_{M}\right).
 \eq
Because of the projectability condition $N = N(t)$, the
Hamiltonian constraint takes  a nonlocal integral form. If one
relaxes projectability and allows $N = N\left(t, x^{i}\right)$,
then the corresponding variation with respect to $N$ will yield a
local super-Hamiltonian constraint ${\cal{L}}_{K} +
{\cal{L}}_{{V}} = 8\pi GJ^{t}$. As argued in \cite{LP}, this will
result in an inconsistent theory. Results obtained by relaxing
projectability should be treated with caution.

Variation with respect to the shift $N^{i}$ yields the
super-momentum constraint,
 \bq \lb{eq2}
\nabla_{j}\pi^{ij} = 8\pi G J^{i},
 \eq
where the super-momentum $\pi^{ij} $ and matter current $J^{i}$
are
 \bqn \lb{eq2a}
\pi^{ij} &\equiv& \frac{\delta{\cal{L}}_{K}}{\delta\dot{g}_{ij}}
 = - K^{ij} + \left(1 - \xi\right) K g^{ij},\nb\\
J^{i} &\equiv& - N\frac{\delta{\cal{L}}_{M}}{\delta N_{i}}.
 \eqn
Varying with respect to $g_{ij}$, on the other hand, leads to the
dynamical equations,
 \bqn \lb{eq3}
&&
\frac{1}{N\sqrt{g}}\left(\sqrt{g}\pi^{ij}\right)^{\displaystyle{\cdot}}
= -2\left(K^{2}\right)^{ij}+2 \left(1 - \xi\right)K K^{ij}
\nb\\
& &~~ + \frac{1}{N}\nabla_{k}\left[N^k \pi^{ij}-2\pi^{k(i}N^{j)}\right] \nb\\
& &~~ + \frac{1}{2} {\cal{L}}_{K}g^{ij}   + F^{ij} + 8\pi G
\tau^{ij},
 \eqn
where $\left(K^{2}\right)^{ij} \equiv K^{il}K_{l}^{j},\; f_{(ij)}
\equiv \left(f_{ij} + f_{ji}\right)/2$, and
 \bqn
\lb{eq3a} F^{ij} &\equiv&
\frac{1}{\sqrt{g}}\frac{\delta\left(-\sqrt{g}
{\cal{L}}_{V}\right)}{\delta{g}_{ij}}
 = \sum^{8}_{s=0}{g_{s} \zeta^{n_{s}}
 \left(F_{s}\right)^{ij} }.
 \eqn
The constants are given by $g_{0} = {2\Lambda}{\zeta^{-2}}$,
$g_{1} = -1$, and $n_{s} =(2, 0, -2, -2, -4, -4, -4, -4,-4)$. The
stress 3-tensor is defined as
 \bq \label{tau}
\tau^{ij} = {2\over \sqrt{g}}{\delta \left(\sqrt{g}
 {\cal{L}}_{M}\right)\over \delta{g}_{ij}},
 \eq
and the geometric 3-tensors $ \left(F_{s}\right)_{ij}$ are defined
as follows:
  \bqn \lb{eq3b}
\left(F_{0}\right)_{ij} &=& - \frac{1}{2}g_{ij},\nb\\
\left(F_{1}\right)_{ij} &=& R_{ij}- \frac{1}{2}Rg_{ij},\nb\\
\left(F_{2}\right)_{ij} &=& 2\left(R_{ij} -
\nabla_{i}\nabla_{j}\right)R
-  \frac{1}{2}g_{ij} \left(R - 4\nabla^{2}\right)R,\nb\\
\left(F_{3}\right)_{ij} &=& \nabla^{2}R_{ij} - \left(\nabla_{i}
\nabla_{j} - 3R_{ij}\right)R - 4\left(R^{2}\right)_{ij}\nb\\
& & +  \frac{1}{2}g_{ij}\left( 3 R_{kl}R^{kl} + \nabla^{2}R
- 2R^{2}\right),\nb\\
\left(F_{4}\right)_{ij} &=& 3 \left(R_{ij} -
\nabla_{i}\nabla_{j}\right)R^{2}
 -  \frac{1}{2}g_{ij}\left(R  - 6 \nabla^{2}\right)R^{2},\nb\\
 \left(F_{5}\right)_{ij} &=&  \left(R_{ij} + \nabla_{i}\nabla_{j}
 \right) \left(R_{kl}R^{kl}\right)
 + 2R\left(R^{2}\right)_{ij} \nb\\
& & + \nabla^{2}\left(RR_{ij}\right) - \nabla^{k}\left[\nabla_{i}
\left(RR_{jk}\right) +\nabla_{j}\left(RR_{ik}\right)\right]\nb\\
& &  -  \frac{1}{2}g_{ij}\left[\left(R - 2 \nabla^{2}\right)
\left(R_{kl}R^{kl}\right)\right.\nb\\
& & \left.
- 2\nabla_{k}\nabla_{l}\left(RR^{kl}\right)\right],\nb\\
\left(F_{6}\right)_{ij} &=&  3\left(R^{3}\right)_{ij}  +
\frac{3}{2}
\left[\nabla^{2}\left(R^{2}\right)_{ij} \right.\nb\\
 & & \left.
 - \nabla^{k}\left(\nabla_{i}\left(R^{2}\right)_{jk} + \nabla_{j}
 \left(R^{2}\right)_{ik}\right)\right]\nb\\
 & &    -  \frac{1}{2}g_{ij}\left[R^{k}_{l}R^{l}_{m}R^{m}_{k} -
 3\nabla_{k}\nabla_{l}\left(R^{2}\right)^{kl}\right],\nb\\
 \left(F_{7}\right)_{ij} &=&  2 \nabla_{i}\nabla_{j}
 \left(\nabla^{2}R\right) - 2\left(\nabla^{2}R\right)R_{ij}\nb\\
 & &    + \left(\nabla_{i}R\right)\left(\nabla_{j}R\right)
  -  \frac{1}{2}g_{ij}\left[\left(\nabla{R}\right)^{2} +
  4 \nabla^{4}R\right],\nb\\
\left(F_{8}\right)_{ij} &=&  \nabla^{4}R_{ij} -
\nabla_{k}\left(\nabla_{i}\nabla^{2} R^{k}_{j}
                            + \nabla_{j}\nabla^{2} R^{k}_{i}
                            \right)\nb\\
& & - \left(\nabla_{i}R^{k}_{l}\right)
\left(\nabla_{j}R^{l}_{k}\right)
       - 2 \left(\nabla^{k}R^{l}_{i}\right) \left(\nabla_{k}R_{jl}
       \right)\nb\\
& &    -  \frac{1}{2}g_{ij}\left[\left(\nabla_{k}R_{lm}\right)^{2}
        -
        2\left(\nabla_{k}\nabla_{l}\nabla^{2}R^{kl}\right)\right].
 \eqn

The matter quantities $(J^{t}, \; J^{i},\; \tau^{ij})$ satisfy the
conservation laws \cite{CNPS},
 \bqn \lb{eq4a} & &
 \int d^{3}x \sqrt{g} { \left[ \dot{g}_{kl}\tau^{kl} -
 \frac{1}{\sqrt{g}}\left(\sqrt{g}J^{t}\right)^{\displaystyle{\cdot}}
  \right.  }   \nb\\
 & &  \left.  \;\;\;\;\;\;\;\;\;\;\;\;\;\;\;\;\;\; +  \frac{2N_{k}}
 {N\sqrt{g}}\left(\sqrt{g}J^{k}\right)^{\displaystyle{\cdot}}
 \right] = 0,\\
\lb{eq4b} & & \nabla^{k}\tau_{ik} -
\frac{1}{N\sqrt{g}}\left(\sqrt{g}J_{i}
\right)^{\displaystyle{\cdot}} - \frac{N_{i}}{N}\nabla_{k}J^{k} \nb\\
& & \;\;\;\;\;\;\;\;\;\;\; - \frac{J^{k}}{N}\left(\nabla_{k}N_{i}
- \nabla_{i}N_{k}\right) =
 0.
\eqn

\section{Cosmological background}

\renewcommand{\theequation}{3.\arabic{equation}} \setcounter{equation}{0}

The homogeneous and isotropic universe is described by the FRW
metric, $ ds^{2} = - dt^{2} + a^{2}(t)\gamma_{ij}dx^{i}dx^{j}$
where $\gamma_{ij}={\left(1 + \frac{1}{4}kr^{2}\right)^{-2}}
{\delta_{ij}}$, with $k = 0, \pm 1$. For this metric, $\bar K_{ij}
= - a^{2}H \gamma_{ij}$ and $\bar R_{ij} = 2k\gamma_{ij}$, where
$H = \dot{a}/a$ and an overbar denotes a background quantity. Then
we find that
 \bqn \lb{3.3}
\bar{\cal{L}}_{K} &=&  3\left(3\xi-2\right) H^{2},\nb\\
\bar{\cal{L}}_{V} &=& 2\Lambda - \frac{6k}{a^{2}} +
\frac{12\beta_1k^{2}}{a^{4}}  + \frac{24\beta_2k^{3}}{a^{6}},
 \eqn
where $\beta_1= {\zeta^{-2}}\left(3g_{2} + g_{3}\right)$ and
$\beta_2 = {\zeta^{-4}}\left(9g_{4} + 3g_{5} + g_{6}\right)$.

Because of the spatial homogeneity, both $\bar{\cal{L}}_{K}$ and
$\bar{\cal{L}}_{V}$ are independent of the spatial coordinates,
and the matter quantities are
 \bq\label{mq}
\bar{J}^t=-2\bar\rho,~~ \bar{J}^i=0,~~ \bar \tau_{ij} = \bar p\,
\bar g_{ij},
 \eq
where $\bar\rho$ and $\bar p$ are the total density and pressure.
Then the Hamiltonian constraint (\ref{eq1}) reduces to the
super-Hamiltonian constraint, $\bar{\cal{L}}_{K}(t) +
\bar{\cal{L}}_{V}(t) =  8\pi G\bar J^{t}(t)$, which leads to the
modified Friedmann equation,
 \bqn \lb{3.4a}
\left(1 - \frac{3}{2}\xi\right)H^{2} + \frac{k}{a^{2}} &=&
\frac{8\pi G}{3}\bar \rho+ \frac{\Lambda}{3}\nb\\  &&~+
\frac{2\beta_1k^{2}}{a^{4}} + \frac{4\beta_2k^{3}}{a^{6}}.
 \eqn
From Eqs.~(\ref{eq2a}) and (\ref{eq3a}) we find that
 \bqn \lb{3.5}
\bar F^{ij} &=& \left(-\Lambda + \frac{k}{a^{2}} +
\frac{2\beta_{1}k^{2}}{a^{4}}  + \frac{12\beta_{2}k^{3}}{a^{6}}
\right)\bar{g}^{ij},\nb\\
 \bar\pi ^{ij}  &=& \left( 3\xi-2\right)H\, \bar g^{ij}.
 \eqn
Then the dynamical equation (\ref{eq3}) reduces to \cite{SVW}
 \bqn \lb{3.4b}
\left(2 - 3\xi\right)\frac{\ddot{a}}{a} &=&  - {8\pi G\over
3}(\bar\rho+3\bar p)+ {2\over3} \Lambda
\nb\\
  & &~~~
- \frac{4\beta_{1}k^{2}}{a^{4}}  - \frac{16\beta_{2}k^{3}}{a^{6}}.
 \eqn
Similarly to general relativity, the super-momentum constraint
(\ref{eq2}) is then satisfied identically, since $\bar J^{i} = 0$
and, from Eq.~(\ref{3.5}),  $\vec\nabla_j \pi^{ij}\equiv
\bar\pi^{ij}{}{}_{|j} = 0$, where $\vec\nabla_i$ denotes the
covariant derivative with respect to $\gamma_{ij}$.

Using Eqs.~(\ref{3.4a}) and (\ref{3.4b}), it follows that in the
background the matter satisfies the same conservation law as in
general relativity,
 \bq \lb{3.4e}
\dot{\bar{\rho}} + 3H \left(\bar\rho +\bar p \right) = 0.
 \eq
This can be also obtained from Eq.~(\ref{eq4a}), while
Eq.~(\ref{eq4b}) is satisfied identically.

In deriving Eq.~(\ref{3.4a}) we followed the usual assumption that
the whole FRW universe is homogeneous and isotropic. In
\cite{Mukb}, it was argued that such an assumption might be too
strong. If one relaxes the assumption and requires that only the
observed patch of our universe is homogeneous and isotropic, one
can introduce the notion of ``dark matter as an integration
constant" of the Hamiltonian constraint~(\ref{eq1}): $\bar
\rho(t)$ in Eqs.~(\ref{3.4a}) and (\ref{3.4e}) can be replaced by
$\bar\rho(t)+ {\cal E}(t)$ in the observable patch, where ${\cal
E}(t)=\mbox{const}/a^3$ in the IR limit
\cite{Mukb,Kobayashi:2009hh}. Beyond the observable patch, ${\cal
E}$ is necessarily inhomogeneous. In order to analyze
perturbations on an FRW background, one needs to restrict the
perturbations to the observable patch, which then raises issues
about matching across the boundary of the observable patch. In our
approach, the background is a homogeneous FRW spacetime, so that
${\cal E}=0$ in the background.

\section{Cosmological  perturbations}

\renewcommand{\theequation}{4.\arabic{equation}} \setcounter{equation}{0}

Linear perturbations of the metric give
 \bqn \lb{4.3a}
\delta{g}_{ij} &=&   a^{2}(\eta)h_{ij}\left(\eta, x^{k}\right)
 ,\nb\\
\delta{N}^{i} &=&   n^{i}\left(\eta, x^{k}\right) ,~ \delta{N} =
a(\eta) n(\eta) ,
 \eqn
where $\eta$ is the conformal time. We decompose into scalar,
vector and tensor modes \cite{MFB92},
 \bqn
\lb{4.4}
n &=& \phi,\;\;\; n_{i} = B_{|i} - S_{i}, \nb\\
h_{ij} &=& - 2\psi\gamma_{ij} + 2E_{|ij} + F_{i|j} + F_{j|i} +
H_{ij},
 \eqn
Note that $\phi$ is a function of $\eta$ only, while $B,\;
S_{i},\; \psi,\; E,\; F$ and $H_{ij}$ are in general functions of
both $\eta$ and $x^{k}$, with the constraints,
 \bqn \lb{4.5}
S_{i}^{\;\;|i} = 0,\;\;F_{i}^{\;\;|i} = 0,\;\;
 H^{i}_{i}  = 0=  H_{ij}^{\;\;\;|j} .
 \eqn

The perturbed energy quantity Eq.~(\ref{eq1a}) is written as
 \bq
\delta J^t=-2\delta \mu.
 \eq
In general relativity, $\delta\mu$ reduces to the density
perturbation $\delta\rho$.

The perturbed matter current in Eq.~(\ref{eq2a}), on the other
hand, decomposes as
 \bq
 \lb{4.20a}
\delta{J}^{i} = \frac{1}{a^{2}}\left( q^{|i} + q^i\right),
~~q^i{}_{|i}=0,
 \eq
and the perturbed stress tensor Eq.~(\ref{tau}) decomposes as
 \bqn \lb{4.25}
&& \delta{\tau}^{ij} = \frac{1}{a^{2}}\Big[\left(\delta{\cal P} +
2\bar{p}\psi\right)\,\gamma^{ij} + {\Pi}^{|\langle ij\rangle}
+ 2\Pi^{(i|j)}
+ \Pi^{ij}\Big], \nb\\
&& \Pi^{i}_{\;\;|i} = 0,~ \Pi^{i}_{i} = 0,\; \Pi^{ij}_{\;\;|j} =
0.
 \eqn
The angled brackets on indices define the trace-free part:
 \bq
f_{|\langle ij \rangle} \equiv f_{|ij}-{1\over 3}
\gamma_{ij}f_{|k}{}^{|k}.
 \eq
In general relativity, $q^{|i}$ and $q^i$ reduce to the scalar and
vector modes of the momentum perturbation $- a(\bar\rho+\bar
p)(v^{|i} + B^{|i} +v^i - S^{i})$, while $\delta{\cal P}$ reduces
to the pressure perturbation $\delta p$, and $\Pi$, $\Pi^i$ and
$\Pi^{ij}$ reduce to the scalar, vector and tensor modes of the
anisotropic pressure.

\subsection{Gauge Transformations}

Consider a gauge transformation as in Eq.~(\ref{gauge}), with
 \bq \lb{4.6}
\chi^{0} = \xi^{0},\;\;\; \chi^{i} = \xi^{|i} +  \xi^{i},\;\;
\xi^{i}_{\;\;|i} = 0,
 \eq
where $\xi^0=\xi^{0}(\eta), \; \xi^{i} =
\xi^{i}\left(\eta,x^{k}\right), \; \xi= \xi
\left(\eta,x^{k}\right)$. Then the metric perturbations in
Eq.~(\ref{4.4}) transform as
 \bqn \lb{4.7a}
&& \tilde{\phi} = \phi - {\cal{H}}\xi^{0} - \xi^{0\prime} ,\;\;
\tilde{\psi} = \psi + {\cal{H}}\xi^{0},\nb\\
&& \tilde{B} = B + \xi^{0}   - \xi',\;\; \tilde{E} = E - \xi, \nb\\
&& \tilde{S}_{i} = S_{i} +   \xi'_{i},\; \tilde{F}_{i} = F_{i} -
\xi_{i},\; \tilde{H}_{ij}= H_{ij},
 \eqn
where ${\cal{H}}= a'/a$ and a prime denotes
$\partial/\partial\eta$. Note that these gauge transformations are
precisely the standard forms given in GR. The only difference is that
in the HL case, $\phi$ and $\xi^0$ are homogeneous. We can
omit $\xi^0$ from $\tilde B$, since only the gradient
of $\tilde B$ occurs in the metric. However, we are free to maintain the $\xi^0$
term -- and we do this in order that we can use
the standard form of the gauge-invariant Bardeen potentials
$\Phi, \Psi$ -- see Eq.~(\ref{4.9a}) below.
Using the gauge freedom, we can restrict
some of the quantities defined in Eq.~(\ref{4.4}).

\subsubsection*{Synchronous Gauge}

This gauge is defined by
 \bq \lb{4.8}
\tilde{\phi} = 0,\;\;\; \tilde{B} = 0, \;\;\; \tilde{S}_{i} = 0,
 \eq
and from Eqs.~(\ref{4.7a}) we find that
 \bqn
\lb{4.9} && \xi^{0} = \frac{1}{a}\int{a\phi d\eta} +
\frac{C_{0}}{a}, \; \xi_{i} =
 \int{S_{i} d\eta} + C_{i}(x),\nb\\
&& \xi = \int{B d\eta} + \int{{d\eta}\over{a}}\left( \int {a\phi
d\eta}\right) + C(x),
 \eqn
where $C(x)$ and $C_{i}(x)$ are arbitrary functions of $x^{k}$
with $C_{i}^{\;\;|i} = 0$, and $C_{0}$ is an arbitrary constant.
Therefore, as in general relativity, this gauge does not
completely fix all the gauge degrees of freedom. This gauge was
used to study the scalar graviton mode in \cite{SVW}.

\subsubsection*{Quasi-longitudinal Gauge}

In general relativity, the longitudinal gauge is defined by
$\tilde{B} = \tilde{E} = \tilde{F}_{i} = 0$ \cite{MFB92}. However,
due to the projectability condition, we see from Eq.~(\ref{4.7a})
that we cannot set all 3 quantities to zero, although we are still
free to set $ \tilde{E} = 0$ and $\tilde{F}_{i} = 0$. In addition,
using the remaining degree of freedom, we can further set
$\tilde{\phi} = 0$. Thus we can set
 \bq
\tilde{\phi} = 0,\;\; \tilde{E} = 0, \;\; \tilde{F}_{i} = 0,
 \eq
with
 \bqn \lb{4.9q}
\xi^{0} =\frac{1}{a}\int{a\phi d\eta} + \frac{C_{0}}{a},\;\; \xi =
E,\; \xi_{i} = F_{i},
 \eqn
which are unique up to a constant $C_{0}$.  We call this the
quasi-longitudinal gauge (it has been used by \cite{Cosmos,Cai} in
the case where projectability
is abandoned, and $k=0$).\\

It should be noted that, as in general relativity, in each of
these gauges only two scalars are left, and  we can define the
same gauge-invariant potentials as in general relativity
\cite{MFB92}
 \bqn \lb{4.9a}
\Phi &=& \phi + {\cal{H}}\left(B - E'\right) + \left(B -
E'\right)',\nb\\
\Psi  &=& \psi - {\cal{H}} \left(B - E'\right).
 \eqn
(Note that in \cite{Cai} a different set of gauge-invariant
variables was used.)

\subsection{Scalar Perturbations\\ in Quasi-longitudinal Gauge}

In the quasi-longitudinal gauge, the metric scalar perturbations
are given by
 \bqn \lb{4.10}
ds^{2} &=& a^{2}\left[- d\eta^{2} + 2 {B}_{|i}dx^{i}d\eta +
\left(1 - 2 \psi\right)\gamma_{ij}dx^{i}dx^{j}\right].\nb\\
 \eqn
Then from Eqs.~(\ref{2.5}) and (\ref{2.6}), we find that
 \bqn \lb{4.13}
K_{ij} &=& \bar K_{ij}   +   a\left[B_{|ij}
 +  \left(\psi' + 2{\cal{H}}\psi\right)\gamma_{ij}\right],\nb\\
 K &=&  \bar K + { }{a^{-1}}
 \left(\vec\nabla ^{2}B + 3 \psi'\right),\nb\\
 K^{ij} &=& \bar K^{ij}   + { }{a^{-3}}\left[B^{|ij}
 +  \left(\psi' - 2{\cal{H}}\psi\right)\gamma^{ij}\right],\nb\\
 {\cal{L}}_{K} &=&  \bar{\cal{L}}_{K} +
 {2 {\cal{H}}}{a^{-2}} \left(2 - 3\xi\right)
 \left(\vec\nabla ^{2}B  + 3 \psi'\right),\; \;\;\;\;\;
 \eqn
and
 \bqn
 \lb{4.15}
 {\cal{L}}_{V} &=&  \bar{\cal{L}}_{V}
  - \frac{4 }{a^{2}}\left(1 - \frac{4\beta_1k}{a^{2}}\right)
  \left(\vec\nabla ^{2} + 3k\right)\psi \nb\\
 & &~ +  \frac{48 \beta_2k^{2}}{a^{6}} \left(\vec\nabla ^{2}
+ 3k\right)\psi  \nb\\ &&~  +  \frac{24  g_{7}
k}{\zeta^{4}a^{6}}\vec\nabla ^{2}
 \left(\vec\nabla ^{2} + 3k\right)\psi.
 \eqn
To first-order the Hamiltonian constraint (\ref{eq1}) is
 \bq
\lb{4.17} \int d^{3}x\sqrt{\gamma}\left(\delta{\cal{L}}_{K} +
\delta{\cal{L}}_{V}\right) = -16\pi G \int{d^{3}x \sqrt{\gamma}\,
\delta{\mu} }.
 \eq
Using Eqs.~(\ref{4.13}) and (\ref{4.15}) we find that
 \bqn \lb{4.18}
& & \int \sqrt{\gamma}d^{3}x\Bigg[\left(\vec\nabla^2+3k\right)\psi
- \frac{(2-3\xi){\cal H}}{2}
\left(\vec\nabla^2 B + 3\psi'\right) \nb\\
& &~~~ -2 k\Big(\frac{2\beta_{1}}{a^{2}} +
\frac{6\beta_{2}k}{a^{4}}
+ \frac{3g_{7}}{\zeta^{4}a^{4}}\vec\nabla^2\Big)\left(\vec\nabla^2+3k\right)\psi\nb\\
& &   ~~~ -{4\pi G a^{2}}\delta{\mu}\Big]=0.
 \eqn
The integrand is a generalization of the general relativity
Poisson equation \cite{MFB92}. Note that the Laplacian terms can
be dropped from this equation, using the identity,
  \bq
\int{d^{3}x\sqrt{\gamma}\, \vec\nabla ^{2}f} = 0.
  \eq

At first-order the supermomentum constraint (\ref{eq2}) is
 \bq \lb{4.20}
\left[ \left(2 - 3\xi\right){\psi}' - 2kB - {\xi}
 \vec\nabla^{2}B\right]_{|i} = 8\pi G a\, {q}_{|i} \,,
 \eq
which generalizes the general relativity $0i$ constraint
\cite{MFB92}. Note that in the general relativity limit ($\xi=0$)
and in a Minkowski background  ($ q = 0 = k$), Eq.~(\ref{4.20})
implies
 \bq \lb{4.20e}
\psi = G(x),
 \eq
where we used the Hamiltonian constraint (\ref{4.18}) to set a
homogeneous function of integration to zero. This result is
closely related to the fact that the spin-0 scalar mode of the
graviton becomes stabilized in the limit $\xi = 0$, as we show in
the next section.

The perturbed dynamical equations require the perturbed
$\left(F_{s}\right)_{ij}$ of Eq.~(\ref{eq3b}). The results are
given by Eq.~(\ref{Fterms}) in the Appendix. Using
Eqs.~(\ref{4.23}) and (\ref{4.25}) in Eq.~(\ref{eq3}), we can find
the perturbed dynamical equations. The trace part gives
 \bqn
\lb{4.24}
 \psi'' &+& 2{\cal{H}}\psi'  - {\cal{F}}\psi - \frac{1}{3(2-3\xi)}
 \gamma^{ij}\delta{F}_{ij}\nb\\
 &+& {1\over 3}\left(\vec\nabla^2B'+2 {\cal H}
\vec\nabla^2B \right) = { 8\pi Ga^{2}\over (2-3\xi)} \delta{\cal
P}.
 \eqn
Here $\delta F_{ij}=\sum g_s \zeta^{n_s}\delta (F_s)_{ij}$, with
$\delta(F_s)_{ij}$ given by Eq.~(\ref{Fterms}),  and ${\cal{F}}$ is
defined as
 \bq
 {\cal{F}} ={2a^2\over(2-3\xi)}\left(-\Lambda+{k \over a^2}+{2\beta_1k^2
 \over a^4} +{12\beta_2 k^3 \over a^6 } \right).
 \eq
The trace-free part is
 \bqn\label{4.24b}
B'_{|\langle ij \rangle}+2{\cal H}B_{|\langle ij \rangle} +\delta
F_{\langle ij \rangle}=-8\pi G a^{2} \Pi_{|\langle ij \rangle}.
 \eqn
These two equations generalize the general relativity $ij$
perturbed field equations \cite{MFB92}.

The perturbed parts of the conservation laws (\ref{eq4a}) and
(\ref{eq4b}) give
 \bqn \lb{4.26a}
& & \int \sqrt{\gamma} d^{3}x \Big[\delta\mu' + 3{\cal H}
\left(\delta{\cal P} + \delta\mu\right)
-3 \left(\bar\rho + \bar p\right){\psi}' \Big] =
0,\nb\\\lb{4.26b}\\ & & \left[q'+3 {\cal H}q  - a\delta{\cal P} -
{2a\over3}\left( \vec\nabla ^{2}+3k \right)\Pi\right]_{|i} = 0. ~~
 \eqn
The energy conservation equation is an integrated generalization
of the general relativity energy equation, and the momentum
equation generalizes the general relativity momentum equation
\cite{MFB92}.

\section{Scalar Graviton on Minkowski background}
\renewcommand{\theequation}{5.\arabic{equation}} \setcounter{equation}{0}

In the general relativity limit on a Minkowski background, the
scalar graviton mode should be suppressed. Otherwise the recovery
of general relativity would be obstructed. By contrast, when $\xi
\neq 0$, we expect the scalar mode may play a significant role.

We set $a = 1$,  $k = 0=\Lambda $,  $J^{t} = 0=J^i$ and
$\tau^{ij}$ = 0. Then
 \bqn
\lb{5.1} \delta{F}_{ij} & =&  - \left(1 + \alpha_{1}\vec\nabla
^{2}  + \alpha_{2}\vec\nabla ^{4}\right)
\left(\psi_{,ij} - \delta_{ij}\vec\nabla ^{2}\psi\right),\nb\\
\delta{\cal{L}}_{K} & =& 0, \;\;\; \delta{\cal{L}}_{V} = -
4\vec\nabla ^{2}\psi,
 \eqn
where $\alpha_{1} \equiv\zeta^{-2}( {8g_{2} + 3g_{3}})$,
$\alpha_{2} \equiv - {\zeta^{-4}}({8g_{7} - 3g_{8}})$. Note that
$\psi=\Psi$ since ${\cal H}=0$, so that $\psi$ is gauge invariant.
The Hamiltonian constraint (\ref{4.17}) is satisfied identically.
(It is interesting to note that if the projectability condition is
given up, the Hamiltonian constraint becomes the super-Hamiltonian
constraint $\delta{\cal{L}}_{K} + \delta{\cal{L}}_{V} = 0$, which
gives the strong condition $\vec\nabla ^{2}\psi = 0$.)

The super-momentum constraint (\ref{4.20}) gives
 \bq \lb{5.4}
\left(2-3\xi\right)\dot{\psi} -{\xi}\vec\nabla ^{2}B = 0.
 \eq
Substituting Eq.~(\ref{5.1}) into the dynamical equations
(\ref{4.24}) with $f = 0$, we find
 \bqn \lb{5.6}
&& \left(2-3\xi\right)\left(3\dot{\psi} + \vec\nabla
^{2}{B}\right)^{\displaystyle{\cdot}} = 2\left(1 +
\alpha_{1}\vec\nabla ^{2}  + \alpha_{2}\vec\nabla ^{4}\right)
\vec\nabla ^{2}\psi.\nb\\
 \eqn

We consider the cases $\xi \not=0$ and $\xi = 0$ (general
relativity limit) separately.

\subsection{$\xi \not= 0$}

From Eqs.~(\ref{5.4}) and (\ref{5.6}) we obtain the wave equation
 \bq \lb{5.7}
\ddot{\psi} - {c_{\psi}^{2}} \left(1 + \alpha_{1}\vec\nabla^{2} +
\alpha_{2}\vec\nabla ^{4}\right) \vec\nabla ^{2} \psi = 0,
 \eq
where $c_{\psi}^{2} \equiv   \xi/(2-3\xi)$. In Fourier space,
 \bqn \lb{5.9a}
\ddot{\psi}_{n} + \omega^{2}_{n} \psi_{n} = 0, ~~\omega^{2}_{n}
\equiv n^{2}c_\psi^{2}\left(1 -  \alpha_1 n^2 + \alpha_2
n^4\right),
 \eqn
where $n$ is the wave-number. It is clear that the solution is
stable in the IR limit, provided that $ 0 \le \xi \le 2/3$, which
is equivalent to $ 1/3 \le \lambda \le 1$, where $\lambda = 1 -
\xi$ is the parameter used in \cite{Horava}. This is in contrast
to the conclusions obtained in \cite{BS}, in which it was found
that $\psi$ is not stable for any choice of $\xi$. The main reason
is that in \cite{BS} the authors considered the case with detailed
balance, or at most with `soft' breaking (and not full breaking)
of detailed balance. With the most general breaking of detailed
balance \cite{SVW}, we find that to have stability in the IR, it
is necessary that $0 \le \xi \le 2/3$ (or $ 1/3 \le \lambda \le
1$).

In addition, $\psi$ is also stable in the UV regime for
$\alpha_{2} > 0$. It is stable in intermediate regimes provided
that either (a)~$\alpha_2>0$ and $\alpha_1\leq 0$ or
$\alpha_{1}^{2} <4 \alpha_{2}$; or (b)~$\alpha_2=0$ and
$\alpha_1\leq 0$.


\subsection{$\xi  = 0$}

When $\xi = 0$, from Eq.~(\ref{4.20e}) we find that $\psi(t,x) =
G(x)$. Inserting this into Eq.~(\ref{5.6}) and integrating,
 \bq \lb{5.12}
\vec\nabla ^{2}B = H(x)  + t \left(1 + \alpha_{1}\vec\nabla ^{2} +
\alpha_{2}\vec\nabla ^{4}\right) \vec\nabla ^{2}G(x),
 \eq
where $H(x)$ is an arbitrary integration function.
Thus it appears that the scalar graviton has a growing mode
$\propto t$. However, it is important to note that $B$ is not
gauge-invariant and therefore $B$ does not directly determine the
stability of the spin-0 scalar graviton mode. The gauge-invariant
variables defined by Eq.~(\ref{4.9a}) are in this case
 \bq \lb{5.12b}
\Phi = \dot{B}=I(x), \;\;\;\;\; \Psi =\psi= G(x).
 \eq
Here $I(x)$ is determined by Eq.~(\ref{5.12}): in Fourier space,
$I_n=(1-\alpha_1 n^2 + \alpha_2 n^4)G_n $.

Clearly, neither of the gauge-invariant variables is growing with
time. As a result, the spin-0 scalar graviton is indeed stable in
the general relativity limit ($\xi = 0$) on a Minkowski
background.

This conclusion appears to contradict the one obtained by SVW
\cite{SVW}. However, a closer analysis shows that in terms of
gauge-invariant variables, the results are consistent. The
synchronous gauge variables used in \cite{SVW} are $\psi$ and
 \bq
\lb{5.12d} h = - 6 \psi +  \vec\nabla ^{2}E,~~ B=0,
 \eq
and they find that
 \bqn \lb{5.12e}
E  = L(x) + M( x) t + Q( x)t^{2}.
 \eqn
The scalar mode appears to be growing because $h$ is. However, by
Eq.~(\ref{4.9a}), the gauge-invariant variables for the SVW
solution are $\Phi=-\ddot E=-2Q(x) $ and $\Psi=\psi= G(x)$ --
neither of which is growing.

Our conclusion is also consistent with the results obtained
recently by Mukohyama \cite{Mukb}.

It is interesting to note that the coupling of the spin-0 scalar
graviton to a dust fluid on a Minkowski background does not alter
this conclusion. In fact, one can show that $\psi$ and $B$ will
satisfy the same equations as above in both cases, $\xi \not= 0$
and $\xi = 0$. The only difference is that now the Hamiltonian
constraint (\ref{4.18}) requires the matter energy quantity to
satisfy the condition $ \int d^{3}{x}\, \delta{\mu} = 0.$

\section{Scalar Perturbations of the Flat FRW Model}
\renewcommand{\theequation}{6.\arabic{equation}} \setcounter{equation}{0}

We return now to an FRW background. In the flat case, $k = 0$, we
find that the perturbation equations simplify considerably.

The super-momentum constraint (\ref{4.20}) reduces to
  \bq \lb{6.2}
  (2-3\xi){\psi}' = {\xi}\vec\nabla^{2}B + 8\pi Ga q.
 \eq
Integrating it over space and using the  Hamiltonian constraint
(\ref{4.18}), we find
 \bq \lb{6.1}
\int{ d^{3}x \left(3 {\cal H}q + a\delta{\mu}\right)} = 0.
  \eq

When $k=0$, we also have
 \bqn \lb{6.4}
&& \delta{F}_{ij} =  2\Lambda a^{2}\psi \gamma_{ij} \nb\\
  &&~~ -  \left(1 +\frac{\alpha_{1}}{a^{2}}\vec\nabla^{2}
+  \frac{\alpha_{2}}{a^{4}}\vec\nabla^{4}\right)
\left(\vec\nabla_{i}\vec\nabla_{j}   -
\delta_{ij}\vec\nabla^{2}\right)\psi. ~~~~~~
 \eqn
Then  the trace-free   dynamical equation (\ref{4.24b}) gives
 \bq \label{6.6}
 \left(a^{2}B\right)'  = \left(a^{2} + \alpha_{1} \vec\nabla ^{2}
+  \frac{\alpha_{2}}{a^{2}}\vec\nabla ^{4}\right) \psi  -8\pi
Ga^{4} \Pi,~~~~
 \eq
while  the trace  equation (\ref{4.24}) reduces to
 \bqn \lb{6.5}
{\psi}'' &+& 2{\cal H}{\psi}' - {\xi \over 2-3\xi}\left(1+
\frac{\alpha_1}{a^2}\vec\nabla^2 +\frac{\alpha_2}{a^4}\vec\nabla^4
\right) \vec\nabla^2\psi
\nb\\
&=&  {8\pi Ga^{2} \over 3(2-3\xi)}\Big[3\delta{\cal P} + (2-3\xi)
\vec\nabla^{2}\Pi\Big].
 \eqn
The conservation laws Eqs. (\ref{4.26a}) and (\ref{4.26b}) reduce
to
 \bqn \lb{6.7a}
& & \int d^{3}x \Big[\delta\mu' + 3{\cal H} \left(\delta{\cal P} +
\delta\mu\right)  -3 \left(\bar\rho +
\bar p\right){\psi}' \Big] = 0,\;\;\;~~\\
\lb{6.7b} & &   q'+3{\cal H}q     = a\delta{\cal P} +
{2\over3}a\vec\nabla^{2}\Pi. ~~~~
 \eqn
Note that not all the equations are independent.
Equation~(\ref{6.5}) can be derived from Eqs. (\ref{6.2}),
(\ref{6.6}) and (\ref{6.7b}). Therefore, we are left with three
first-order evolution equations, (\ref{6.2}), (\ref{6.6}) and
(\ref{6.7b}), and two integral constraints, Eqs. (\ref{6.1}) and
(\ref{6.7a}), for the six unknowns, $\psi, B, \delta{\mu},
\delta{\cal{P}}, q$ and $\Pi$.

In terms of the gauge-invariant variables defined in
Eq.~(\ref{4.9a}), we can rewrite Eq.~(\ref{6.6}) as
  \bq
  \lb{6.8}
\Phi - \Psi =   -8\pi Ga^{2} \Pi+ \frac{1}{a^{2}}\left(\alpha_{1}
+ \frac{\alpha_{2}}{a^{2}}\vec\nabla ^{2}\right)  \vec\nabla
^{2}\psi.
 \eq
The last term on the right acts as an effective anisotropic stress
from HL gravity, i.e. from the higher-order curvature terms:
 \bq
\Pi_{\text{grav}}=-{1 \over 8\pi G a^4}\left(\alpha_{1} +
\frac{\alpha_{2}}{a^{2}}\vec\nabla ^{2}\right)  \vec\nabla
^{2}\psi.
 \eq
%
This stress is strongest on small scales, and is suppressed on
large scales. It might provide a signal to distinguish the HL
theory from general relativity.

\section{Conclusions}

In this paper, we systematically studied the linear scalar
perturbations of the FRW models in the SVW setup \cite{SVW}, which
is the most general theory of HL type \cite{Horava} when detailed
balance is abandoned, but projectability is maintained (and so is
parity).

We generalized \cite{SVW}, who considered only a vacuum Minkowski
background. In addition to generalizing the geometrical terms, we
included matter and derived the conservation laws. We have not
specified the type of matter, as it is still an open question how
to construct the matter Lagrangian ${\cal{L}}_{M}$, although
scalar and vector fields have recently been studied
\cite{KK,calcagni}.

Working in the quasi-longitudinal gauge, we obtained explicitly
the perturbed Hamiltonian constraint (\ref{4.18}), the
super-momentum constraint (\ref{4.20}), and the  dynamical
equations (\ref{4.24}) and (\ref{4.24b}). The perturbed
conservation laws are given by Eqs.~(\ref{4.26a}) and
(\ref{4.26b}).

A crucial issue in the HL theory and its generalizations is the
spin-0 scalar graviton mode. By specializing the FRW background to
its Minkowski limit, we showed via a gauge-invariant treatment
that this mode is stable in the IR limit for $0 \le \xi \le 2/3$.
It is also stable in the UV regime, provided that the arbitrary
coupling constants $g_{7}$ and $g_{8}$ are suitably chosen. The
apparent contradiction with the results of \cite{SVW} is resolved
via a gauge-invariant reformulation of their results. This is
consistent with the results of Mukohyama \cite{Mukb}. We also
showed that this conclusion is true when coupling the scalar
graviton to a dust fluid in Minkowski spacetime. This result is
also different from  the one obtained in \cite{BS}, in which it
was shown that the scalar mode is not stable for any given $\xi$.
The main reason is that in \cite{BS} the authors considered the
case with detailed balance, or at most 'soft' breaking of detailed
balance.

The stability condition $0 \le \xi \le 2/3$ has the unwanted
consequence that the scalar mode is a ghost
\cite{Horava,Mukb,BPS,KA}. To tackle this problem, one may
consider the theory in the range $\xi <0$ and then try to remove
the instability of the scalar mode via the Vainshtein mechanism
\cite{Vain}.

Our general formulas for the FRW background provide the basis for
further work to analyze cosmological tensor perturbations,
inflationary perturbations and large-scale structure formation in
the framework of the generalized HL theory. We showed that there
is an effective gravitational contribution to the anisotropic
stress on small scales, Eq.~(\ref{6.8}), so that in HL theory we
have $ \Phi \neq \Psi$ even in the absence of matter anisotropic
stresses.

~\\{\bf Acknowledgements:} We thank R.-G. Cai, D. Coule, K.
Koyama, H. L\"u, D. Matravers, S. Mukohyama, A. Papazoglou, M.
Sasaki, S. Seahra, Y.-S. Song,  T. Sotiriou, M. Visser and D.
Wands  for valuable discussions and suggestions. AW thanks the
Institute of Cosmology and Gravitation for their hospitality. AW
was partially supported by Baylor University and the NSFC grant,
No. 10703005 and No. 10775119. RM was supported by the UK's
Science \& Technology Facilities Council.

\appendix
\section{Perturbed $\left(F_{s}\right)_{ij}$}
\renewcommand{\theequation}{A.\arabic{equation}} \setcounter{equation}{0}

To first-order, the $\left(F_{s}\right)_{ij}$ are given by \bqn
\lb{Fterms} \left(F_{0}\right)_{ij} &=& - \frac{1}{2}a^{2}
\gamma_{ij} +
a^{2}\psi\gamma_{ij},\nb\\
\left(F_{1}\right)_{ij} &=& -k \gamma_{ij} +   \left[\psi_{|ij} -
\left(\vec{\nabla}^{2}\psi\right)\gamma_{ij}\right],\nb\\
%
\left(F_{2}\right)_{ij} &=& \frac{6k^{2}}{a^{2}} \gamma_{ij} +
\frac{4  k}{a^{2}}\left[3\psi_{|ij}
+ \gamma_{ij}\left(\vec{\nabla}^{2} + 3k\right)\psi\right]\nb\\
& & -  \frac{8  }{a^{2}}\left(\vec{\nabla}_{i} \vec{\nabla}_{j} -
\gamma_{ij}\vec{\nabla}^{2}\right)\left(\vec{\nabla}^{2}
+ 3k\right)\psi,\nb\\
%
\left(F_{3}\right)_{ij} &=& \frac{2k^{2}}{a^{2}} \gamma_{ij} +
\frac{2  k}{a^{2}}\gamma_{ij}\left(\vec{\nabla}^{2} + 2k\right)\psi\nb\\
& & -   \frac{ 1}{a^{2}}\left(3\vec{\nabla}^{2} +
10k\right)\left(\psi_{|ij}
- \gamma_{ij}\vec{\nabla}^{2}\psi\right)\nb\\
& & + \frac{4
}{a^{2}}\left[\vec{\nabla}^{2}\left(\psi_{|ij}\right)
- \left(\vec{\nabla}^{2}\psi\right)_{|ij}\right],\nb\\
%
\left(F_{4}\right)_{ij} &=& \frac{108k^{3}}{a^{4}} \gamma_{ij}
+ \frac{36  k^{2}}{a^{4}}\left[3\psi_{|ij} 
+ \gamma_{ij}\left(5\vec{\nabla}^{2} + 12k\right)\psi\right]\nb\\
& & - \frac{144  k}{a^{4}}\left(\vec{\nabla}_{i} \vec{\nabla}_{j}
- \gamma_{ij}\vec{\nabla}^{2}\right)\left(\vec{\nabla}^{2} + 3k\right)\psi,\nb\\
%
\left(F_{5}\right)_{ij} &=& \frac{36k^{3}}{a^{4}} \gamma_{ij} +
\frac{24  k^{2}}{a^{4}}\left[\psi_{|ij} 
+ 2\gamma_{ij}\left(\vec{\nabla}^{2} + 3k\right)\psi\right]\nb\\
& & - \frac{2
k}{a^{4}}\left[3\vec{\nabla}^{2}\left(\vec{\nabla}_{i}
\vec{\nabla}_{j} -3\gamma_{ij}\vec{\nabla}^{2}\right)\psi  \right.\nb\\
& &  \;\;\;\;\; \;\;\;\;\;\;\;\;  +  6\left(\vec{\nabla}_{i}
\vec{\nabla}_{j}
-3\gamma_{ij}\vec{\nabla}^{2}\right)\vec{\nabla}^{2}
\psi\nb\\
& &  \;\;\;\;\; \;\;\;\;\;\;\;\;   \left.
     + 2\gamma_{ij}\left(\vec{\nabla}^{2} -37k\right)\vec{\nabla}^{2}
     \psi\right],\nb\\
%
\left(F_{6}\right)_{ij} &=& \frac{12k^{3}}{a^{4}} \gamma_{ij} +
\frac{12  k^{2}}{a^{4}}\left(\vec{\nabla}^{2} +
4k \right)\psi  \gamma_{ij} \nb\\
& & - \frac{6
k}{a^{4}}\left[2\left(\vec{\nabla}^{2}\psi\right)_{|ij}
 + \vec{\nabla}^{2}\left(\psi_{|ij}\right) \right.\nb\\
& & \left.  \;\;\;\;\; \;\;\;\;\;\;\;\; -
\gamma_{ij}\vec{\nabla}^{2}
\left(3\vec{\nabla}^{2} + 8k\right)\psi\right],\nb\\
\left(F_{7}\right)_{ij} &=& \frac{8
}{a^{4}}\left\{\left[\left(\vec{\nabla}^{2} +
3k\right)\vec{\nabla}^{2}\psi\right]_{|ij}
       \right.\nb\\
 & & \left. - \gamma_{ij} \left(\vec{\nabla}^{2} + 3k\right)
 \left(\vec{\nabla}^{2} + 2k\right) \vec{\nabla}^{2}\psi\right\},\nb\\
\left(F_{8}\right)_{ij} &=&     \frac{
1}{a^{4}}\left\{\vec{\nabla}^{4}\left(\psi_{|ij}\right) -
\vec{\nabla}^{k} \vec{\nabla}_{i}\vec{\nabla}^{2}\left(\psi_{|jk}
\right)\right.\nb\\
& & - \vec{\nabla}^{k}
\vec{\nabla}_{j}\vec{\nabla}^{2}\left(\psi_{|ik}\right) - 2
\vec{\nabla}_{i} \vec{\nabla}_{j}\vec{\nabla}^{2}
\left(\vec{\nabla}^{2} + 4k\right)\psi\nb\\
& &   + 2\gamma_{ij}\vec{\nabla}^{4}\left(\vec{\nabla}^{2} +
4k\right)\psi\nb\\
& & \left. + \gamma_{ij}\vec{\nabla}^{k}
\vec{\nabla}^{l}\vec{\nabla}^{2}\left(\psi_{|kl}\right)\right\},
\eqn where $ \vec{\nabla}^{k}\psi \equiv \psi^{|k}$. In addition,
we also have
 \bqn \lb{4.23}
& & \frac{1}{N\sqrt{g}}\left(\sqrt{g}\pi^{ij}\right)' =
- \left(2 - 3\xi\right)\frac{\ddot{a}}{a^{3}}\gamma^{ij}\nb\\
& & \;\;\;\;\;\;\; + \frac{ 1}{a^{3}}\left\{ \left(2 -
3\xi\right)\gamma^{ij}\left(a\ddot{\psi} + 2\dot{a}\dot{\psi}
-2\ddot{a}\psi\right)\right.\nb\\
& & \;\;\;\;\;\;\; \left.- \left[\dot{B}^{|ij} -
\left(1-\xi\right)
\gamma^{ij}\vec\nabla ^{2}\dot{B}\right]\right\},\nb\\
& & \left(K^{2}\right)^{ij} -  \left(1-\xi\right) KK^{ij}
 =  \left( 3\xi-2\right)\frac{H^{2}}{a^{2}}\gamma^{ij}\nb\\
& & \;\;\;\;\;\;\;  +  \frac{ {H}}{a^{3}}\left\{\left(1 -
3\xi\right)B^{|ij} + \left(1 - \xi\right)\gamma^{ij} \vec\nabla
^{2}
{B}\right.\nb\\
& & \;\;\;\;\;\;\; \left. + 2\left(2 -
3\xi\right)a\left(\dot{\psi}
- H\psi\right) \gamma^{ij}\right\},\nb\\
&& {\cal{L}}_{K} g^{ij} = - \frac{3(2-3\xi)H^{2}}{a^{2}}
\gamma^{ij}\nb\\
& & \;\;\;\;\;\;\;  +  \frac{ 2(2-3\xi){H}}{a^{3}}
\left[\vec\nabla ^{2}B  + 3a\left(\dot{\psi} - H\psi\right)\right]
 \gamma^{ij},\nb\\
& & N^k\nabla_k \pi^{ij}+2\pi^{k(i} \nabla_kN^{j)}
 + \pi^{ij}\nabla_{k}N^{k}\nb\\&&\;\;\;\;\;\;\; =
 \frac{ (2-3\xi){H}}{a^{2}}\left(2B^{|ij}
 -  \gamma^{ij} \vec\nabla ^{2}B\right).
 \eqn


\end{document}